\documentstyle[prl,aps,preprint]{revtex}

\begin{document}
\draft

\title{Phase transition in inelastic disks}
\author{Teruhisa S. Komatsu}
\address{Department of Pure and Applied Sciences,
 University of Tokyo, Komaba, Meguro-ku, Tokyo 153, Japan}

\maketitle
\begin{abstract}
This letter investigates the molecular dynamics of inelastic disks
 without external forcing.
By introducing a new observation frame with a rescaled time,
 we observe the virtual steady states converted from
 asymptotic energy dissipation processes.
System behavior in the thermodynamic limit is carefully investigated.
It is found that a phase transition with symmetry breaking occurs
 when the magnitude of dissipation is greater than a critical value.
\end{abstract}


In energy conservative systems,
 it is well known that
 macroscopic properties at an equilibrium state
 can be described by a few state variables \cite{Callen}.
This raise a question as to whether, in dissipative systems,
 similar state variables might be defined for macroscopic properties
 at steady states maintained by an external energy source.
When a dissipation rate is sufficiently small, the system
 might be expected
 to behave as a conservative system; but is this actually the case?
To examine this and other related questions, it is
 interesting to investigate models
 that can connect a conservative and a dissipative system
 by varying the system parameters.

An ensemble of elastic hard disks is one of the simplest models 
 for describing the fluid state in conservative systems.
Computer simulations played an important role 
 in discovering the existence of fluid-solid transition in this model
 \cite{AlderWainwright62}.
We will here consider an ensemble of inelastic hard disks,
 which has previously been investigated as a model of
 granular materials\cite{granular}.
By varying the inelasticity, we can continuously
 change the system from a conservative to a dissipative system.
This ability will be useful for our present purposes of investigating
 a thermodynamic properties in dissipative systems.

In order to attain a steady state in a dissipative system,
 an external energy source that compensates for energy dissipation
 due to inelastic collisions is indispensable.
In granular physics, the vibrating bed \cite{vib} is a typical energy source.
Nevertheless, attaching such an energy source
 will break the isotropy of the system from the outset.
In this letter, we simulate the dissipative system
 without energy sources
 and investigate the virtual steady states using a new observation frame
 that will be introduced later.

The inelastic disk system without energy input
 , known as a cooling gas or dissipative gas system,
 has been investigated using molecular dynamics (MD)
 and hydrodynamic models.
These studies reveal that the homogeneous state is unstable
 for sufficiently high inelasticity
 even without attaching any energy source.
In this state,
 a cluster of particles \cite{GoldhirschZanetti93}
 and collective mean flow (shear mode)\cite{McNamaraYoung94}
 are formed.
We study how the instability is characterized
 by means of MD simulation of an ensemble of inelastic hard disks,
 with a particular focus on the asymptotic collective phenomena
 and its system size dependence.


The system consists of an ensemble of $N$ inelastic hard disks
 in two dimensional space.
For simplicity, each particle is assumed to have a unit mass and
its rotational degrees of freedom are ignored.
Collisions between circular particles are inelastic;
the inelastic collision is implemented in the following manner.
At a collision of two particles,
the tangential velocities to the collision plane are preserved,
while the normal component of the relative velocity $\Delta v_n$
changes to $\Delta v'_n$, where
\begin{equation}
  \Delta v'_n=-e_n\Delta v_n .
\label{eq:collisionrule}
\end{equation}
Here, the coefficient of restitution $e_n (0\le e_n \le 1)$
is constant for all collisions.
In the following, we parameterize the inelasticity
by $\epsilon (\equiv 1-e_n)$.
In the case $\epsilon=0$, the system becomes conservative
with the elastic collisions; in the case $\epsilon>0$,
energy dissipation occurs due to the inelastic collisions.

The selected boundary conditions consist of
 a square box enclosed by elastic rigid walls.
Thus the energy of the system is not dissipated
 by the bouncing of particles against the wall.
For initial conditions, we adopt spatially homogeneous states
 equilibrated by setting $\epsilon=0$.
The time evolution of the system is calculated by
 the event-driven method\cite{AlderWainwright}.

In this letter, we focus on asymptotic behavior of the system 
 after a sufficiently long transient time.
We note that our model is almost identical with
 the model in \cite{McNamaraYoung94} except for boundary conditions.

Under our boundary conditions,
 there are three important system parameters:
 the number of particles $N$,
 the inelasticity $\epsilon$
 and the area fraction of particles $\phi$.
Here $\phi (0\le\phi<1)$ is the ratio of
 the total area covered by particles to that of the box $S$.

In the conservative case $\epsilon=0$, it is known that this system,
 consisting of an ensemble of elastic hard disks, has two phases:
 a fluid and a solid phase.
In this letter, we investigate
 the system in the range of parameter $\phi$
 corresponding to the fluid phase when $\epsilon=0$.
For this parameter range, our results do not depend on
 the dispersity of particle radii $a$.
In the following, we show only the results
 for monodisperse case($a=0.5$);
 however, the results of our simulations for the polydisperse case 
 (uniform distribution in the range $0.4\le a\le 0.5$)
 are almost the same.
On the other hand, in the range of parameter $\phi$
 corresponding to the solid phase when $\epsilon=0$, 
 the results depend on the dispersity of particle radii.
The results for this parameter range will be
 reported in a future publication.

For sufficiently large values of $\epsilon$,
 it is known that the time development by the event driven method
 is ill-defined because infinite collisions occur
 in a finite time interval
 (inelastic collapse) \cite{McNamaraYoung94}.
Our investigation is limited to sufficiently small values of $\epsilon$ 
 in order to avoid the inelastic collapse.

We will now describe how to observe the system.
Since no energy source is attached to the system,
the total energy of the system monotonically decreases in time.
Nevertheless, by {\it the rescaling of time},
 it is possible to make the energy of the system virtually conserved.
We introduce the rescaled time $\tilde{t}$ and
  the rescaled velocity of $i$-th particle $\tilde{v}_i$:
\begin{equation}
\tilde{t}=\int_0^t \sqrt{T(t)} dt,\quad\tilde{v}_i=v_i(t)/\sqrt{T(t)} ,
\label{eq:scaletv}
\end{equation}
where $T(t)$ is the averaged kinetic energy per one particle
 at a time $t$, $T(t)=\sum_i v^2_i(t)/2N$;
 and $v_i(t)$ is the original velocity of $i$-th particle.
We call the rescaled system ``R-system''
 and the original system ``O-system''.
All of the present observations are carried out for the ``R-system''.
Under the translation by Eqs.(\ref{eq:scaletv}),
 the averaged kinetic energy per one particle 
 in the ``R-system'' is normalized to the unity,
 $\tilde{T}(t)\equiv\sum_i\tilde{v}_i^2(t)/2N=1$,
 i.e., the total kinetic energy is conserved
 for any $\epsilon$ in the ``R-system''.
Further, asymptotic energy dissipation processes
 in the ``O-system'' are translated to steady states in the ``R-system''.

It must be noted here that the translation by Eqs.(\ref{eq:scaletv})
 is just the rescaling of time and that
 the trajectories of the particles in space
 are not influenced by this rescaling.
We calculate the time evolution of the ``O-system''
 by the event driven method,
 and observe the ``R-system'' with a special focus on its steady states
 after a sufficiently long transient time.
In the rest of this letter,
 all variables shown refer to those in the ``R-system''.


The dependencies of pressure on $\phi$ for several values
 of $\epsilon$ are shown in Fig.\ref{fig:pressure}.
Pressure $P$ is defined as the time averaged sum of the impulses
 at the bouncing of particles on the walls
 per unit length per unit time in the ``R-system''.
The vertical axis in the figure is $N\tilde{T}/PS$,
 which is unity for the ideal gas limit ($\phi\rightarrow 0$).
Here $S$ is the area of the box.

In the conservative case $\epsilon=0$,
 only the fluid phase exists in the system for $\phi<\phi_c$.
The value of $\phi_c$ is reported to be
 $\phi_c\simeq 0.7$ \cite{AlderWainwright62,Ito}.
Consider the dissipative cases $\epsilon>0$
 in Fig.\ref{fig:pressure}(a)
 paying attention to the difference from the case $\epsilon=0$.
When $\epsilon=0.02$, there is no remarkable difference.
For $\epsilon=0.04$, a decrease in pressure is observed
 in the intermediate range of $\phi$.
For $\epsilon=0.08$, a similar difference appears
 in the wider range of $\phi$.

Comparing Fig.\ref{fig:pressure}(a) and (b),
 it is found that similar dependencies of pressure on $\phi$
 are observed for the same values of $N\epsilon$.
Thus $N\epsilon$ is an important parameter of the system and
 it would characterize the distance from the conservative system.

For sufficiently large $N\epsilon$,
 the behavior of the system is quite different
 from the case $\epsilon=0$.
A snapshot of the system for $(N,\epsilon)=(256,0.08)$
 is shown in Fig.\ref{fig:meanflow}(b) compared to
 that for $\epsilon=0$ (Fig.\ref{fig:meanflow}(a)).

From Fig.\ref{fig:meanflow}(b),
 it is found that a correlation of particle velocity exists and that
 a mean flow circulating anti-clockwise in the box is formed.
As a result of the emergence of the circulating mean flow,
 the impulse of particles to the walls at bounce
 decreases compared with the case $\epsilon=0$.
Then the decrease of pressure, i.e., the increase of
 $N\tilde{T}/PS$, occurs.
For larger values of $\epsilon$,
 the circulation is formed in a wider range of $\phi$,
 as seen in Fig.\ref{fig:pressure}.

Since the circulation is formed above
 a certain threshold value of $\epsilon$,
 normalized total angular momentum $M$ of the system
 is defined as an order parameter of the system.
\begin{equation}
M = \frac{1}{N \sqrt{S}} \sum_{i=1}^N (r_i-R_0) \wedge \tilde{v}_i .
\end{equation}
Here $r_i$ and $\tilde{v}_i$ are coordinates
 and velocities of $i$-th particle,
$R_0$ are the coordinates of the center of the box,
``$\wedge$'' refers to outer product,
and the notation $\tilde{v}$ is used
 to remind us that we are observing the ``R-system''.
$M$ is normalized by the box length $\sqrt{S}$ to eliminate 
 the dependency on the box size.

Time developments of $M$ for some values of $\epsilon$
 are shown in Fig.\ref{fig:Mtseq}, where $N=256$ and $\phi=0.5$.
For $\epsilon=0$, the value of $M$ fluctuates around $M=0$.
As $\epsilon$ is increased, the fluctuations of $M$ increase.
For $\epsilon=0.036,0.038$, the circulating mean flow is formed.
The direction of the circulation sometimes turns over.
As $\epsilon$ increases further, the events of turning over become rare.
In order to investigate the formation of the circulation in detail,
 the distributions of $M$ are shown in Fig.\ref{fig:Mdist};
 because these distributions are symmetric around $M=0$,
 only the region $M\ge 0$ is shown.

As denoted by Fig.\ref{fig:Mdist}, it is found
 that the peak position of $M$, $M_{\rm peak}$
 becomes nonzero for $\epsilon$ above a certain threshold value
 and increases continuously from zero with increasing $\epsilon$.
We also observe a similar behavior
 when we vary the value of $\phi$
 while fixing the value of $\epsilon$.
Thus the circulation appears continuously for any direction
 in the parameter space($\epsilon,\phi$).
In Fig.\ref{fig:Mdist}, the distributions
 of $M$ are broad in shape,
 which indicates that the value of $M$ fluctuates
 around the value at the peak $M_{\rm peak}$.
If the formation of the circulation is
 {\it a phase transition} accompanied by symmetry breaking
 in the thermodynamic limit ($N\rightarrow\infty$),
 the width of the distribution will converge to zero in this limit.
In order to confirm this scenario,
 the distributions of $M$ are examined in the dependence on $N$
 in Fig.\ref{fig:MdistN}, where $N$ is
 varied while $N\epsilon$ is kept constant.

In Fig.\ref{fig:MdistN}(a), it is found that $M_{\rm peak}$
 will converge in the limit, $N\rightarrow\infty$.
This convergence shows that $M_{\rm peak}$ is
 a function of $N\epsilon$ for sufficiently large $N$,
 because the value of $N\epsilon$ is fixed
 in Fig.\ref{fig:MdistN}(a).

From Fig.\ref{fig:MdistN}(b), it is found that
 the width of the fluctuation of $M$
 decreases in the manner of $1/\sqrt{N}$ as $N$ increases.
Thus, we conclude that the circulation
 appears as a phase transition with symmetry breaking.
The turning over events of the circulation
 observed in Fig.\ref{fig:Mtseq} are finite size effects.

In Fig.\ref{fig:MvsNeps}, 
 $M_{\rm peak}$ around the critical point is shown.
The horizontal axis is $N\epsilon(1-3/\sqrt{N})$,
 where the second term $O(1/\sqrt{N})$ comes from finite size effects.
This figure clearly shows that $M_{\rm peak}$ is a function of $N\epsilon$
 for sufficiently large $N$.
Further calculations are needed
 for the precise determination of the critical exponents.


In this letter, we simulate inelastic disks in a square box
 and investigate the steady states of the system
 using a new observation frame
 wherein the energy of the system is virtually conserved by
 {\it the rescaling of time}.
The important parameter of the system is $N\epsilon$, where
 $N$ is the number of particles and $\epsilon$ is inelasticity.
At $\epsilon=0$, the steady states are homogeneous
 equilibrium states.
At sufficiently small $N\epsilon$, the steady states
 are also homogeneous.
For $N\epsilon$ above the critical points,
 the homogeneous steady states
 are no longer stable \cite{CommGZ}.
Then the phase transition
 with symmetry breaking occurs
 where the circulation appears continuously.

The results shown in this letter are independent of
 the characteristics of the model.
As noted before, the results are independent of
 the dispersity of the particle radii.
The hard-core potential is not essential to the results.
We have confirmed that the similar results are obtained
 in the case of soft-core potential: a simulation of the models using
 the discrete element method\cite{DEM}
 with Nose-Hoover thermostat\cite{NoseHoover}
 without the rescaling of time.
We have also observed similar circulating mean flow in
 the system with inelastic hard spheres.
Thus similar results would also be obtained in three dimensions.

A number of problems remain to be investigated.
For example, for larger $\epsilon$ up to $\epsilon=1$, are there
 one or more additional phases?
Is hydrodynamic description by $N\rightarrow\infty$
 with fixing $N\epsilon$
 really possible for any boundary conditions?
Do fluid- and solid phases exist in $\epsilon\ne 0$?
How are the phenomena in the system investigated here related to
 those in the system with a real energy source, such as vibrating beds?

It is noteworthy that
 the homogeneous states in the system with finite $\epsilon$
 are always unstable for sufficiently large $N$,
 because the phase transition occurs at finite $N\epsilon$.
Thus the conservative system is a singular-limit system
 when we take the limit $N\rightarrow\infty$ first.

The author thanks S. Sasa and N. Nakagawa for their useful input
 and careful reading of the manuscript, N. Ito and Y.-h. Taguchi
 for their helpful advice, 
 and Dept. of Math. Sci. at Ibaraki Univ. for their hospitality.
The author also acknowledges the support from JSPS.

\begin{figure}
\caption{Dependency of pressure $P$ on $\phi$.
The vertical axis is $N\tilde{T}/PS$ and the horizontal
 axis is area fraction $\phi$.
(a) $N=256$ and $\epsilon=0,0.02,0.04,0.08$.
(b) $N=1024$ and $\epsilon=0,0.005,0.01,0.02$.
In the intermediate range of $\phi$ and
 for sufficiently large $\epsilon$,
 differences of the dependency are remarkable.
Comparing (a) with (b), a similar dependence is observed
 for the same values of $N\epsilon$.}
\label{fig:pressure}
\end{figure}

\begin{figure}
\caption{(a) Snapshot of the system for $\epsilon=0$.
 (b) Snapshot of that for $\epsilon=0.08$.
Here, $N=256$,$\phi=0.5$.
The lines from the center of each particle show
 velocities of particles.
Velocity correlation and a circulating mean flow
are clearly observed in (b).}
\label{fig:meanflow}
\end{figure}
\begin{figure}
\caption{Time dependences of $M$. Here, $N=256,\phi=0.5$.
The horizontal axes are time in the ``R-system'' $(0\le\tilde{t}\le 40000)$,
 while the vertical axes are $M$ $(-0.5\le M\le 0.5)$.
In each figure, the value of $\epsilon$ is shown at the right side.
For the parameters shown here, the mean free time is a value
 from $3.4$ to $3.6$.}
\label{fig:Mtseq}
\end{figure}

\begin{figure}
\caption{Distribution of $M$.
The horizontal axes are $M$ $(0\le M\le 0.5)$, where the vertical axes
 are frequency.
The value of $\epsilon$ is shown at the left hand
 side of each figure.
Other parameters are same as in Fig.\protect\ref{fig:Mtseq}.}
\label{fig:Mdist}
\end{figure}

\begin{figure}
\caption{$N$-dependence of the distribution of $M$.
Here, $\phi=0.5$ and $N \epsilon = 20.48$.
The number of particles $N$ are $64$(dotted line)
,$256$(single dotted line),$1024$(broken line), and $4096$(solid line).
(a) The horizontal axis is $M$ and the vertical axis is frequency.
(b) The horizontal axis is the difference from the peak position
 times $\protect\sqrt{N}$.
The vertical axis is frequency which is normalized at the peak value.
}
\label{fig:MdistN}
\end{figure}

\begin{figure}
\caption{$M_{\rm peak}$ around the critical point.
The vertical axis is $M_{\rm peak}$ and 
the horizontal axis is $r=N\epsilon (1-3/\protect\sqrt{N})$.
The guide line is $|r-6.85|^{1/2}/6.5$.
}
\label{fig:MvsNeps}
\end{figure}


\begin{references}
\bibitem{CommGZ}
Notice that a similar system size dependence for
 the clustering instability condition was reported
 in \cite{GoldhirschZanetti93} for the hydrodynamic model.
\end{references}
\end{document}